\documentclass[11pt,twoside]{article}
\usepackage{asp2004}
\usepackage{epsf}
\usepackage{psfig}
\begin{document}

\title {SNLS -- the Supernova Legacy Survey}

\author {C.J. Pritchet, for the SNLS Collaboration\footnotemark}
\footnotetext{See {\tt http://snls.in2p3.fr/people/snls-members.html} for a complete list of SNLS collaboration members.}
\affil{Dept. of Physics and Astronomy, U. Victoria, PO Box 3055, Victoria, BC V8W 3P6, Canada}

\begin{abstract}

Type Ia supernovae (SNeIa) provide direct evidence for an
accelerating universe, and for the existence of ``dark
energy" driving this expansion. The Supernova Legacy Survey
(SNLS) will deliver many hundreds of SNIa
detections, and well-sampled $g'r'i'z'$ light curves, over the next 5
years. Using these data, we will obtain a precise measurement of the
cosmological parameters ($\Omega_{mass}, \Omega_{\Lambda}$); our goal is to
determine the cosmological equation of state parameter $w$ to a
precision better than $\pm$0.10, and hence test theories for the
origin of the universal acceleration.

SNLS uses the CFHT MegaCam imager (400 Megapixels, 1 deg$^2$) to image
four fields around the sky in 4 filters, with typical time
sampling of 3--4 nights. A total of 202 nights of CFHT time has been
allocated over the next 5 years for these observations; a large
program of followup spectroscopy is now underway at VLT, Gemini, Keck,
and Magellan.

SNLS has been running since August 2003.  There now exist about 330 
reliable SN detections with excellent light curves out to beyond
redshift 0.9, of which about 80  have been spectroscopically
identified as Type Ia's. See {\tt http://www.cfht.hawaii.edu/SNLS} for
up-to-the-minute information on the latest SN discoveries.

\end{abstract}

\noindent

\section {Introduction}

In late 1998 two teams (Riess et al. 1998, Perlmutter et al. 1999)
independently announced that the expansion of the Universe is
accelerating. This remarkable discovery, which was made using
observations of Type Ia supernovae, ranks as one of the most exciting
developments in cosmology over the past 80 years. We now know from
these, and other, observations that the geometry of the Universe is
exquisitely flat; two-thirds of its energy content consists of a
mysterious component known as ``Dark Energy", which drives the
universal acceleration, and whose density decreases slowly, or not at
all, as the Universe expands.

The key parameter in studying dark energy is the equation of state
parameter $w$, which relates the pressure and the density of the
Universe (through $w=P/\rho$).  For instance, a classical fixed
cosmological constant, $\Lambda$, as proposed by Einstein, yields
$w=-1$, whereas other models (e.g. quintessence) yield values of $w >
-1$ (e.g. Huterer and Turner 2001).

The primary goal of the Supernova Legacy Survey (SNLS) is to distinguish
between dark energy models (and hence strongly constrain the physics
that might lead to dark energy) using luminosity distance measurements
of supernovae. To do this requires a concerted program to find, and
measure the properties of, many hundreds of type Ia supernovae at
redshifts $0.2<z<0.9$ (lookback times of billion of years). Such a
sample of supernovae represents an increase by a factor of order ten
in the number of supernovae available for cosmological analysis; the
chosen redshift range optimally spans lookback times that are most
sensitive to the transition from a matter-dominated to a dark
energy-dominated Universe.

SNLS will allow a high confidence
discrimination between w=$-$1, the ``Einstein value", and w=$-$0.8, a value
that is predicted in one of the simplest available quantum gravity
field theories. Type II supernovae (massive star core collapse) will
also reveal the star formation rate of the Universe in the distant
past, and hence critically constrain the evolution of galaxies over
more than half of the age of the Universe.

\section {SNLS -- an Overview}

SNLS\footnote{\tt http://www.cfht.hawaii.edu/SNLS/}
 is built on the Deep survey --the largest single component
of the Canada-France-Hawaii Telescope Legacy Survey\footnote{\tt
http://www.cfht.hawaii.edu/CFHTLS/}; CFHTLS is possible thanks
to the availability of the 1 deg $\times$ 1 deg MegaCam\footnote{\tt
http://www.cfht.hawaii.edu/Instruments/Imaging/MegaPrime/} 
mosaic imager at CFHT. 

There are a number of attributes of SNLS that make
it attractive for studies of supernovae.

\begin{table}[!ht] \begin{center} \caption{SNLS Fields}
\medskip
{\small
\begin{tabular}{cccl}
\tableline
\noalign{\smallskip}
Field & RA(2000) & Dec(2000) & Other Observations\\
\noalign{\smallskip}
\tableline
\noalign{\smallskip}
D1 &  02:26:00.00 & $-$04:30:00.0 & XMM Deep, VIMOS, SWIRE, GALEX  \\
D2 &  10:00:28.60 & +02:12:21.0 & Cosmos/ACS, VIMOS, SIRTF, XMM  \\
D3 &  14:19:28.01 & +52:40:41.0 & Groth strip, Deep2, ACS  \\
D4 &  22:15:31.67 & $-$17:44:05.7 & XMM Deep  \\
\noalign{\smallskip}
\tableline
\end{tabular} }\end{center}\end{table}

\begin{figure}[!ht]
\plottwo{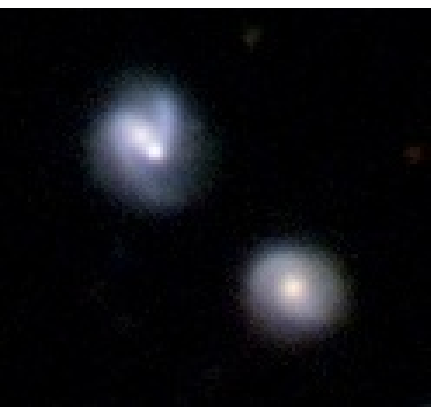}{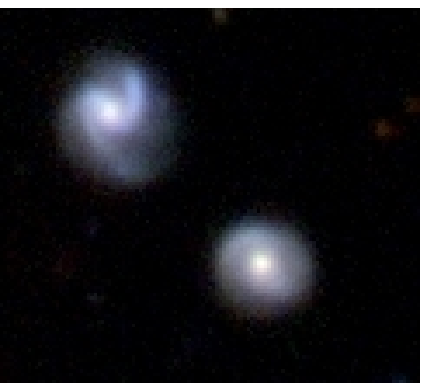}
\caption{A supernova at redshift 0.28
discovered in the SNLS supernova survey.
{\itshape Left:} maximum light; {\itshape right:} after the supernova has faded.}
\end{figure}

{$\triangleright$} {\itshape Sampling and Total Observing Time --}
More than 500 epochs will be obtained over 5 years on each of four 1
square degree fields (for a total of 202 nights of observing
time). (Field coordinates are given in Table 1.)  Typical photometric
sampling of the supernova light curves is once every $3-4$ days
($\sim$2 days in the rest-frame of the supernova) during dark-grey time, 
and with bright time sampling gaps (when MegaCam is taken
off the telescope) typically 11 days or less. 

Each field has a 6
month observing window per year, resulting in extensive light curves
and improved survey efficiency (since a larger fraction of the
supernovae that explode in a 6 month window will be useful for light
curve fitting, compared to shorter observing windows). The time
sampling is a key parameter in the survey design; it greatly improves
the measurement of maximum brightness, and the light curve which is
used to determine the intrinsic luminosity of the supernova. The
observations are obtained in a ``queue-scheduled'' mode by observatory
staff. Without queue scheduling and service observing, the time
sampling and enormous amount of observing would be impossible to
handle.

The field size of MegaCam has a multiplex advantage -- it allows us to
follow the late time light curves of supernovae at the same time that
new discoveries are being recorded.  The sampling cadence permits, for
the first time, a detailed observation of the rising light
curves of a large sample of high redshift SNe.

Figure 1 shows an example of a relatively low redshift (z=0.28)
supernova found in the presurvey phase of observations. Fig. 2
shows some of our light curves of intermediate redshift objects
that reached maximum in Sep--Oct 2003. It can be seen that, even in
the presence of gaps in the light curves (due to bright of moon,
when MegaCam is not mounted, and weather), maximum light can be
measured quite accurately  - typically to an accuracy much better
than $\pm$0.1 mag.

\begin{figure}[!ht]
\plotone{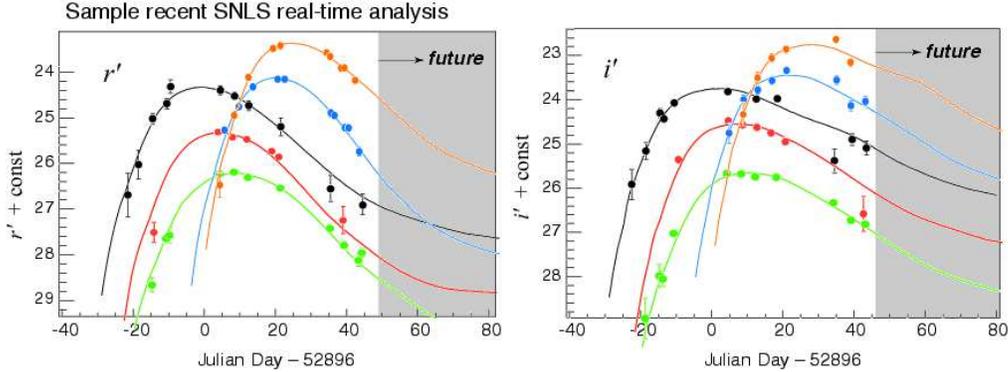}
\caption{Composite light curves (brightness vs. time) from early
SNLS data. Time 0d corresponds to 2003 Sep 14 UT.
It is clear that the maximum brightness, on which
cosmological analysis depends, can be measured to very high accuracy
from these data.}
\end{figure}

{$\triangleright$} {\itshape Filter Coverage -- } SNLS obtains
time-sequenced images in each of the filters $g'r'i'z'$, which are
close to (though not identical with) the Sloan Survey filter set.
Some observations are also taken in $u^*$. These observations permit
excellent corrections from observed (redshifted) wavelengths to the
fiducial restframe $B$ (440nm -- blue light) band, and also allow 
corrections for internal absorption in the host galaxies due
to dust absorption.

{$\triangleright$} {\itshape The Team --} SNLS is a large
international collaboration led by scientists in Canada and France,
with participants in the USA, UK, and elsewhere in Europe. The collaboration
includes scientists at all levels, with 4 scientists in Canada/France
assigned data-handling duties to keep up with the data flow and basic
data reduction. The Chair of the Collaboration Board is Reynald Pain
(LPNHE, U. Paris), and the CFHTLS Supernova Coordinator is Chris
Pritchet (U. Victoria).

{$\triangleright$} {\itshape Data Analysis Pipelines --} Starting
from CFHT real-time preprocessed data (Magnier and Cuillandre 2004), two
independent real-time analysis pipelines (run by the Canadian and French teams)
analyze the data as it is arrives from Mauna Kea at the CFHT
headquarters.  These pipelines produce lists of candidates, and
magnitudes, in about 4--6 hours, and agree quite well down to about
$i'_{AB} = +24$ (redshift about 0.8 for a typical SNIa). The key
element of these pipelines is matching the point spread function of an
exposure to a reference image.  This is done using the Alard (1997)
algorithm for the French team, and using a non-parametric approach
(Pritchet 2004) by the Canadian team.  A complication is the
large ($\sim 1.5$ arcmin) dithering pattern that is used to ``fill
in'' the two 80 arcsec wide gaps in the MegaCam mosaic. This prevents
the use of a chip-by-chip analysis of the mosaic, because much of the
area of each CCD chip would be lost because of the dithering. Instead
it is necessary to ``swarp'' (E. Bertin, private communication) each
individual exposure of a sequence to an astrometric reference frame,
prior to combining and PSF-matching.

Eventually our goal is to merge the two detection pipelines. However,
we plan to maintain two independent paths for photometric analysis,
since this is on a critical path to the derived cosmological model.

\section {Spectroscopic Followup} 

Spectroscopy is vital in order to obtain SN redshifts, and to confirm
the type of each SN.  This requires observations on the world's
largest (8-10 metre class) telescopes, because of the faintness of the
supernovae.  Spectroscopic followup time has been committed for
2003-2005 at the European Southern Observatory Very Large Telescope
(PI Pain), and time is also being used at Gemini North and South (PIs
Hook/Pritchet/Perlmutter), the Keck Observatories (PI Perlmutter, with
complementary spectroscopic followup observations PI'd by Ellis), and
Magellan (PI Carlberg, with complementary IR observations by Magellan
staff). In fact, more spectroscopic 6.5-8-10m telescope time has been
allocated so far than CFHT discovery time!  The organization of this
spectroscopic followup campaign has been one of the major successes of
the SNLS project.

A key element of the spectroscopic followup strategy is the
queue-scheduling and rolling search discovery mode at CFHT. This leads
to improved efficiency for spectroscopic followup because: (1) it
allows us to monitor the rise of the object and trigger spectroscopy
at maximum light; (2) the flux of the target is well known since it is
measured one or two days before max; this allows us an improved
estimate of exposure time; and (3) pre-maximum colours and fluxes
allow good discrimination of SNeIa from other events (e.g. SNeII and
AGN's).

Another success story in the spectroscopic followup is the use of
``nod and shuffle'' observations at Gemini; this mode
virtually eliminates systematic sky residuals for the faintest objects.
See Fig. 3 for an example of this mode of observation.

It is conceivable in the future that some of the spectroscopic typing of
supernovae will be replaced by multi-filter/multi-epoch
typing. The $ugriz$ filter observations of CFHTLS will play a pivotal
role in defining photometric indices that may help to discriminate
SNeIa from other classes of events. We emphasize, however, that
this goal is still far from being realized. 
Furthermore, photometric redshifts do not yet have sufficient precision
to permit their use in Hubble diagram cosmology; spectroscopic (though
non-time-critical) z's for SN hosts will still be required for the
foreseeable future, even if multicolor SN typing were to become practical.

\begin{figure}
\plotone{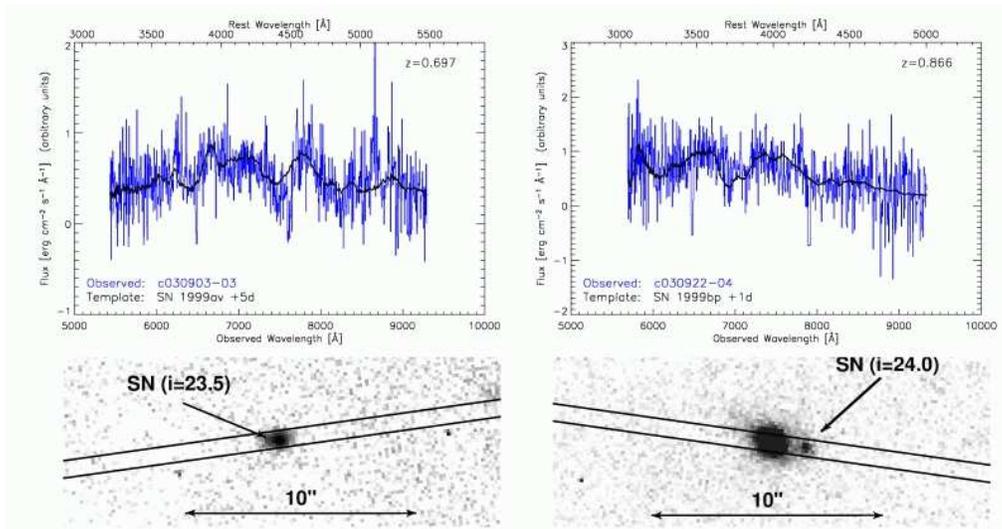}
\caption{Spectra from our Sep 2003 Gemini 8m telescope
observations with the GMOS-N spectrograph,
overplotted with best-fitting local supernova templates. {\itshape Left:} a
SNIa at z=0.697 ($i'=23.5$), observed in classical long slit mode for
3600s; {\itshape right:} an SNIa at z=0.866 ($i'=24.0$), observed in ``nod
and shuffle'' mode for 4800s. Note the much-improved sky-subtraction
for the fainter object observed with nod-and-shuffle (available only
on Gemini).}
\end{figure}

\begin{table}[!ht] \begin{center} \caption{Detections and Real-Time Analysis of Spectroscopy}
\medskip
{\small
\begin{tabular}{lcccc}
\tableline
\noalign{\smallskip}
Run& Detections & Spectra  & SNIa$^*$ & SNII$^*$ \\
\noalign{\smallskip}
\tableline
\noalign{\smallskip}
Pre-survey & 74 & 25 & 15 &  4 \\
2003 Aug & 18 & 10 & 5 & 3 \\
2003 Sep & 33 & 16 & 11 &   - \\
2003 Oct & 28 & 14 & 6 &  2 \\
2003 Nov & 18 & 4 & 3 &  - \\
2003 Dec & 17 & 10 & 6 &  - \\
2004 Jan & 42 & 13 & 6 &  1 \\
2004 Feb & - & - & - & - \\
2004 Mar & 37 & 18 & 8 &  1 \\
2004 Apr & 26 & 19 & 8 &  1 \\
2004 May & 29 & $\dag$ & $\dag$ &  $\dag$ \\
\noalign{\medskip}
All Runs & 322 & 129 & 68 &  12 \\
\noalign{\smallskip}
\tableline
\multicolumn{5}{l}{$*$ Confirmed or probable typing.} \\
\multicolumn{5}{l}{$\dag$ Final spectroscopic statistics not available.} \\
\end{tabular} 
}\end{center}\end{table}

\section {Current Status of SNLS}

The SNLS team is now routinely delivering web-based SN detections and photometry
within 6--12 hr of data being taken. The reader is referred to the
SNLS web pages (see the footnote in \S 2), with links to both the Canadian
and French detection web sites.

Table 2 shows the current (May 2004) status of the  observations.
More than 300 candidate supernovae have been discovered; of these more
than 120 have spectroscopy. A preliminary redshift distribution
of some of the data, with a few explanatory notes, is shown in Fig. 4.
Cumulative statistics of probable SN candidates and spectroscopic
confirmation are shown in Fig. 5.

The weather at Mauna Kea since Oct 2003 has been the worst
in more than 20 years; this resulted in one entire run being lost, 
poor detection statistics, and worse than expected light curve and filter
sampling, through late 2003 and some of 2004. Weather problems have also
affected image quality. However, there is also the underlying problem that
the performance of the MegaCam corrector is not quite as good as expected,
with a fairly noticeable degradation in the corners. The effect of this
is quite difficult to quantify, but early indications are that it has
not affected the numbers of detected supernovae, or the photometry,
significantly.

\begin{figure}[!ht]
\plotone{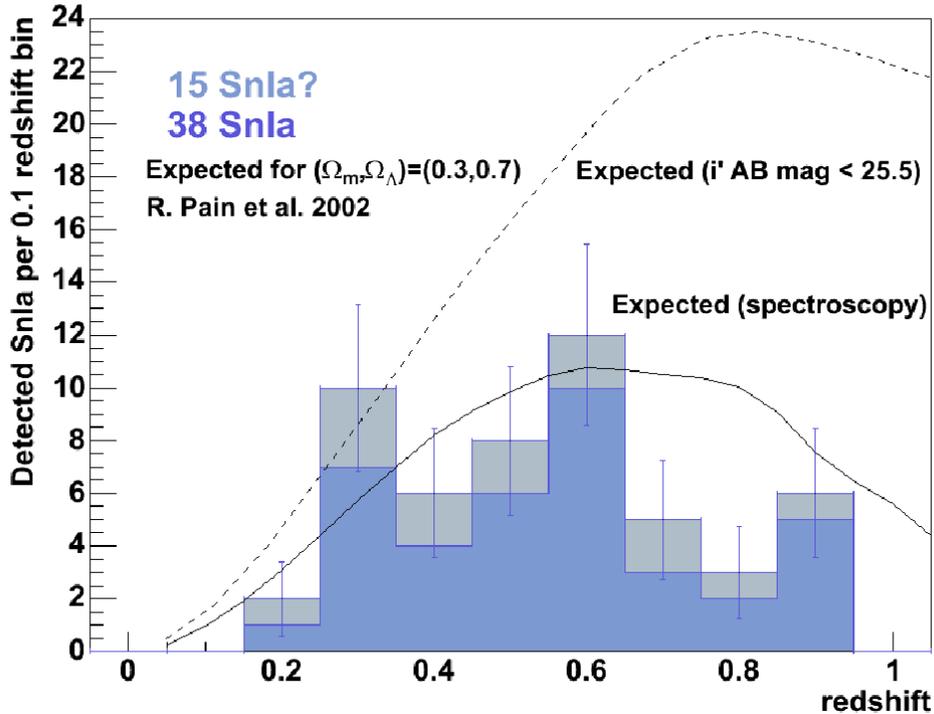}
\caption{\textbf{ Preliminary} numbers of SN Ia and probable SN Ia as a function of
redshift, as of Feb 2004, compared with the expected numbers of Ia
assuming a detection limit of $i'_{AB}$=25.5 (dotted line) and a
spectroscopy limit of $i' =24.5$ (full line). Expected numbers were
computed assuming a flat $\Omega_m=0.33 $ Universe and a distant SNIa
rate from Pain et al. (2002).}
\end{figure}

\begin{figure}[!ht]
\plotfiddle{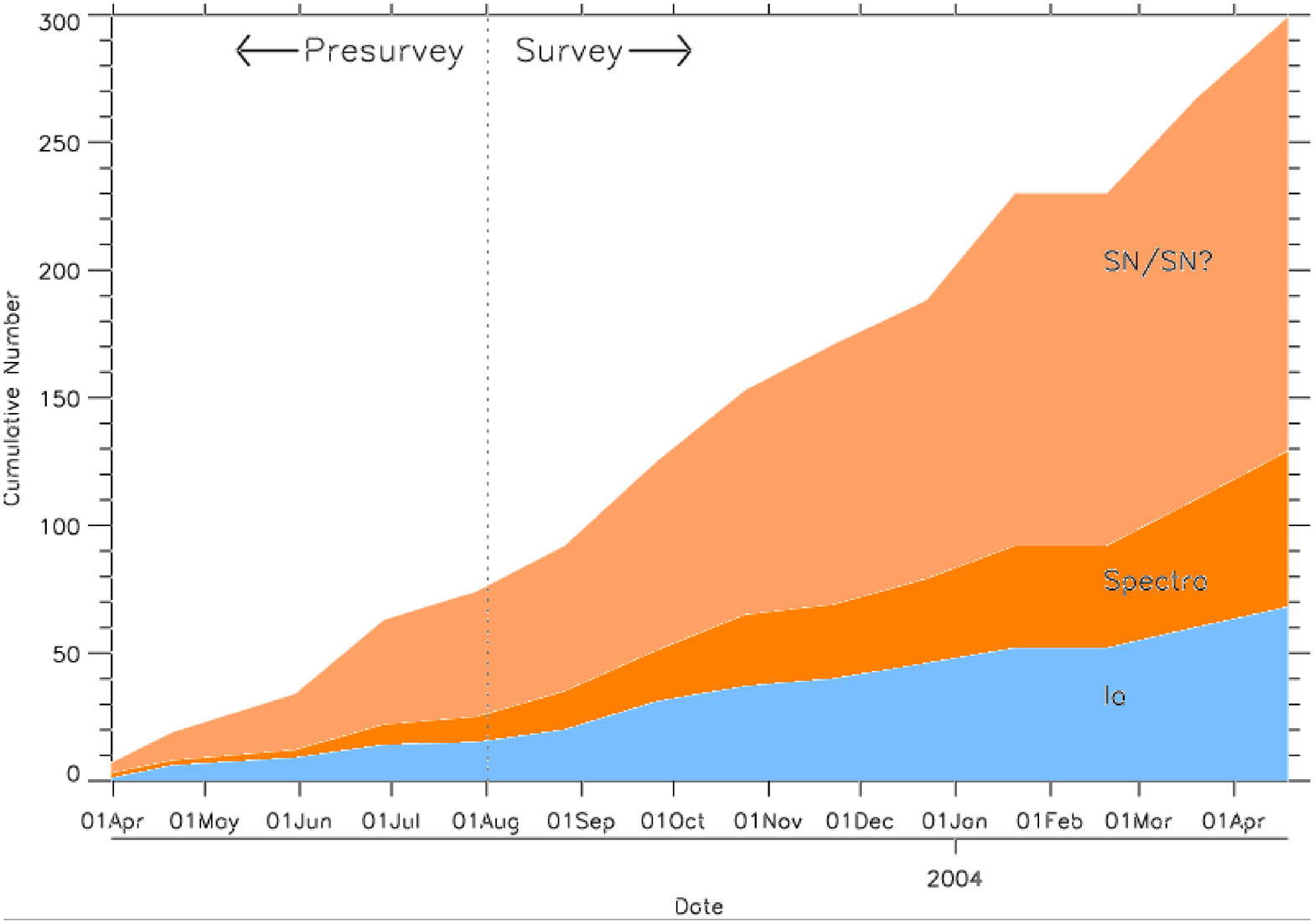}{3.0truein}{0.}{40}{40}{-130}{-50}
\caption {The cumulative numbers of objects discovered since
the start of observing. The upper curve (SN/SN?) shows cumulative
numbers of supernova candidates; the middle curve shows numbers
of objects for which spectroscopy was obtained; and the bottom
curve shows the numbers of SNeIa that were spectroscopically
confirmed. This plot represents a lower limit to our expected
discovery rate.}
\end{figure}

One of the surprising issues to emerge from the Tucson DE meeting was
the widespread concern about calibration. Of course, there are many
``routine'' (!) matters that must be addressed in calibrating a new
instrument such as MegaCam, and work has only recently begun in
earnest on these detailed calibration issues.  Many of these issues
are mitigated by the fact that we are continuously observing the same
fields, and refer each SN to a grid of nearby secondary standard
stars.  We believe we will be able to achieve our goal of $\pm$0.02
mag or better accuracy.

Beyond this, there exist a host of small systematic effects that may
disturb our ability to $\sqrt{N}$ the errors for hundreds of SNe.
These sources of error include: error in the relative absolute
calibration of different filters; variations in color terms over the
field of view (due to, for example, spatially variable wavelength
response of the filters); and detailed (and difficult to determine)
corrections for the convolution of the complex spectral energy
distribution of a SNIa with the response characteristics of the
filter+CCD.

\begin{figure}[!ht]
\plotone{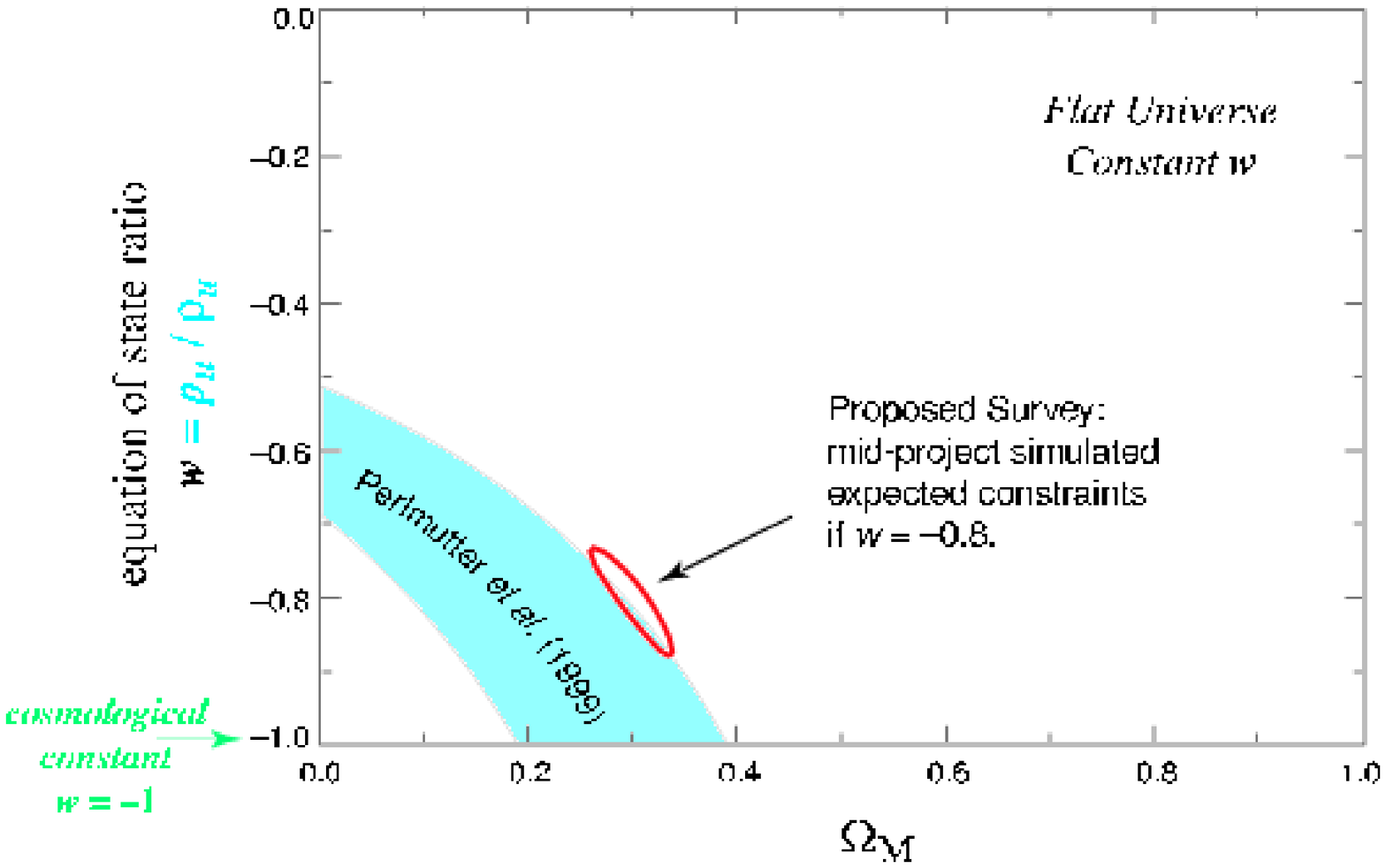}
\caption{Confidence region in the ($w - \Omega_M$) plane,
assuming a flat Universe. The blue region represents the SNe observed
in Perlmutter et al. (1999). The red ellipse is the simulated
1$\sigma$ contour for 300 SNe (SNLS mid-project), assuming that
$\Omega$ is measured independently. The simulation assumes w of --0.8
and $\Omega_M=0.3\pm0.03$. This
 demonstrates the ability to test whether a cosmological constant
fits the data, or whether some other form of dark energy is
required.}
\end{figure}

\section {Science Goals}

The confirmed sample of SNeIa will be used to obtain a precise
measurement of the cosmological parameters $\Omega_M$ and
$\Omega_\Lambda$ (where $\Omega_M$ and $\Omega_\Lambda$ are the
fraction of closure density in matter and vacuum energy; $\Omega_M +
\Omega_\Lambda = 1$ represents a flat Universe). The SNeIa will
be used to obtain a measurement of the dark energy
parameter $w$ with a precision approaching $\pm$0.05 
when a prior on $\Omega_M$ is used -- see Fig. 6.

Supernova observations can be subjected to many straightforward tests
to check for systematic errors. The CFHTLS $u*g'r'i'z'$ data can be used
to measure SN colours and hence test for reddening by comparison with
colours for nearby SNe. (We have collaborative plans with Magellan
staff to use IR data to obtain colour measurements out to $z\sim 0.6$.)
In addition we will have a large enough
sample to be able to study subsets divided by host galaxy type
(derived from the host galaxy spectra or high resolution imaging), or
galactocentric radius - this allows us to check for effects associated
with changes in the underlying host galaxy population (metallicity,
extinction, age), similar to the study of Sullivan et al (2003). 
The SN spectra themselves can be stacked to
obtain a high S/N mean spectrum for different host galaxy types or z
ranges; these can be compared against local SN spectra to check for
small evolutionary effects. Complementary spectroscopic
observations at Keck (Ellis et al.) will also provide a detailed
comparison of low- and high-z SNe.

SNLS is an important step towards a precision measurement of the
dark energy equation-of-state parameter $w$: it will assume constant $w$, and test the possibility that
the dark energy is just the cosmological constant, i.e. the zero-point
energy of the vacuum -- perhaps the simplest, best known dark energy
model. A more exhaustive study sensitive to time-variable $w$ may have
to await the launch of the JDEM mission well into the next decade.

\acknowledgments {SNLS relies on observations with MegaPrime, a joint
project of CFHT, CEA/DAPNIA, and HIA. The SNLS collaboration wishes to
gratefully acknowledge the assistance and co-operation of the CFHT
Queued Service Observing team, headed by Pierre Martin; and
Jean-Charles Cuillandre, Eugene Magnier, Christian Veillet, and Kanoa
Withington for assistance and advice with CFHTLS data, calibration, and CFHT
computer systems. The French collaboration members acknowledge support
from CNRS/IN2P3, CNRS/INSU and CEA. Canadian collaboration members are
funded and supported by NSERC and CIAR. }

\vfill\eject




\begin{references}
\reference Alard, C. 1997, A\&A, 321, 424
\reference Magnier, E.A., \& Cuillandre, J.-C. 2004, PASP, 116, 449
\reference Huterer, D., \& Turner, M.S. 2001, Phys Rev D64, 123527
\reference Pain, R. et al. 2002, ApJ 577, 120
\reference Perlmutter, S., et al. 1999, ApJ, 517, 565
\reference Pritchet, C.J. 2004, in preparation
\reference Riess, A., et al. 1998, AJ, 116, 1009
\reference Sullivan, M., et al. 2003, MNRAS, 340, 1057
\end{references}
\end{document}